\documentclass{aastex62}

\usepackage{wrapfig}
\usepackage[caption=false]{subfig}
\usepackage{subfloat}
\received{}
\revised{}
\accepted{}
\submitjournal{ApJ}

\shorttitle{GRS 1758 \& 1E 1740 with \textit{INTEGRAL}}
\shortauthors{Rodi et al.}
\begin{document}

\title{High Energy Jet Emission from GRS 1758\(-\)258 \& 1E 1740.7\(-\)2942 with \textit{INTEGRAL?}}

\author[0000-0003-2126-5908]{James Rodi}
\affiliation{INAF - Istituto di Astrofisica e Planetologia Spaziali; via Fosso del Cavaliere 100; 00133 Roma, Italy}

\author[0000-0002-2017-4396]{Angela Bazzano}
\affiliation{INAF - Istituto di Astrofisica e Planetologia Spaziali; via Fosso del Cavaliere 100; 00133 Roma, Italy}

\author[0000-0003-0601-0261]{Pietro Ubertini}
\affiliation{INAF - Istituto di Astrofisica e Planetologia Spaziali; via Fosso del Cavaliere 100; 00133 Roma, Italy}



\begin{abstract}

GRS 1758\(-\)258 and 1E 1740.7\(-\)2942 are two long-known persistent black hole binaries in the Galactic Center region.  Using INTEGRAL's extensive monitoring of the Galactic Center and Bulge, we studied their temporal and spectral evolutions in the 30\(-\)610 keV energy range from March 2003 through April 2022 with the IBIS/ISGRI gamma-ray telescope. Our analyses found that the sources typically had Comptonized spectra, though not always with the same parameters.  The spectral states with more than 8 Ms of observation time show deviations from a Comptonized spectrum above \(\sim 200\) keV or a ``hard tail" that extends up to at least 600 keV.  The origin of this component remains debated with the most popular scenarios being synchrotron emission from the jet or Comptonization in a hybrid thermal/non-thermal plasma.  Anyway, the GRS 1758\(-\)258 and 1E 1740.7\(-\)2942 spectra are acceptably described by \texttt{CompTT+po} (jet) and \texttt{Eqpair} (hybrid Comptonization) scenarios.  To differentiate between the two scenarios, we calculated the Spearman correlation coefficient comparing 30\(-\)50 keV count rates with those in higher energy bands (50\(-\)100, 100\(-\)300, and 300\(-\)600 keV).  The count rates below 300 keV are strongly correlated, indicating those photons arise from the same physical process.  Above 300 keV the count rates are either anti-correlated or not correlated with the 30\(-\)50 keV count rates for GRS 1758\(-\)258, which suggests that the photons originate from a different physical process.  For 1E 1740.7\(-\)2942, the level of correlation is unclear due to scatter in the data points.  However, the 300\(-\)600 keV count rates are consistent with a constant value. This disfavors the hybrid Comptonization scenario for both sources.

\end{abstract}



\section{Introduction} \label{sec:intro}

1E 1740.7\(-\)2942 (hereafter 1E) and GRS 1758\(-\)258 (GRS) are two of a handful of persistent, though variable, Black Hole Binaries in the Galaxy. In particular these two sources belong to the Low Mass X-ray Binary (LMXB) class and spend most of their time in the so-called hard state and with thermally dominated soft state part of the time and intermediate states. They both are observed at various energy bands and considered micro-quasars because of their radio characteristics, i.e. radio jets \citep{rodriguez1992,mirabel1999} and double-lobed counterparts \citep{marti2017}.  The sources are located in a crowded field close by to the Galactic Center at a distance of 8 \(\pm\) 1 kpc.  For GRS it was estimated taking into account Jet-ISM interaction \citep{tetarenko2020}. 

The spectral and flux variability for 1E and GRS are similar and, being classified as LMXBs, accrete matter via Roche lobe overflow from the companion star. They are in general in the hard state and show few transitions into intermediate and dim states. For the hard state, spectra typically consist of a power law component with a photon index around 1.6 and exponential cut-off \( < \)50\(- \)100 keV. The soft and intermediate state spectra are characterized by a disk component and a power law with photon index \( > 2\). 

Several previous works have reported on relatively short ISGRI observations of GRS or 1E.  For example, \cite{pottschmidt2006} studied the hard to soft transition in 2003 and 2004 with RXTE/PCA and ISGRI while \cite{delsanto2005} investigated the spectral variability of 1E during the same period also with PCA and ISGRI.  Later analyses of 1E by \cite{castro2014} looked at spectral variability during brief periods in 2003, 2005, and 2012 with XMM-Newton, INTEGRAL/JEM-X, and ISGRI, and \cite{natalucci2014} investigated the spectral and timing properties of 1E in the hard state with ISGRI and NuSTAR observations from 2012.  Hereafter we report about the long-term characteristics of the two sources with high energy broad band results from ISGRI.

\section{Instruments and Observations} \label{sec:inst}

In October 2002 the \textrm{INTErnational Gamma-Ray Astrophysics Laboratory (INTEGRAL)} satellite was launched from Baikonur, Kazachstan \citep{jensen2003}.  The satellite is outside Earth's radiation belts for \(\sim\)85\% of the 2.5\(-\)3 day orbit.  We utilized the IBIS/ISGRI (Imager on Board INTEGRAL/INTEGRAL Soft Gamma-Ray Imager \citep{ubertini2003,lebrun2003}) data in the 30\(-\)900 keV energy range.  

GRS and 1E have been frequently monitored throughout the \textrm{INTEGRAL} mission with observations during Galactic Center and Bulge programs.  The observations used in the subsequent analyses of GRS are within \(10^{\circ}\) of the spacecraft pointing direction during revolutions 46\(-\)2495 (MJD 52699\(-\)59693, 2003-03-01 to 2022-04-25 UTC).  The same angular and temporal constraints were applied in selecting observations of 1E.  This results in 18 422 pointings, or science windows (scw), for GRS and 18 457 scws for 1E resulting in 26.00 Ms of total exposure time for GRS and 24.65 Ms of exposure time for 1E.  

Data reduction was performed using the standard INTEGRAL Offline Science Analysis (OSA) software version 11.2.  The light curve analysis was performed in 5 energy bins from 30\(-\)900 keV while the spectral analysis used 62 energy bins in the same energy range.  A systematic error of 1.5\% was added to the spectral fitting.

%\begin{wrapfigure}{r}{0.5\textwidth}
 \begin{figure}[h!]
  \begin{center}
  \includegraphics[scale=0.9, angle=0,trim = 9mm 15mm 10mm 100mm, clip]{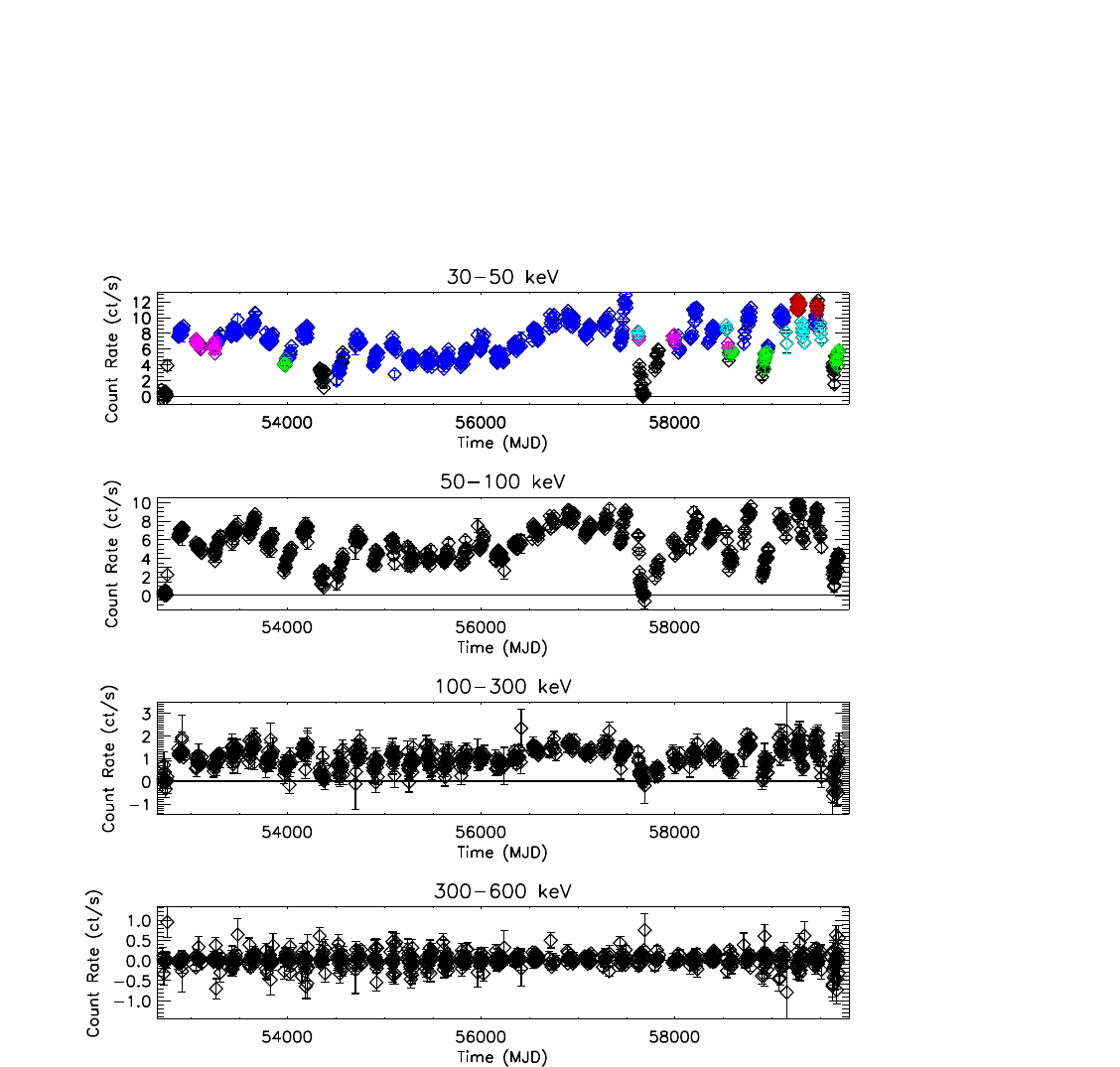}
  \caption{ISGRI long-term light curves of GRS in the 30\(-\)50, 50\(-\)100, 100\(-\)300, and 300\(-\)600 keV energy bands.  Black solid line denotes \(0\) ct/s.  The different colors in the 30\(-\)50 keV light curve correspond to different spectral states discussed in Sec.~3.2.1.} 
  \label{fig:lc_grs}
      \end{center}
%    \hspace{-60mm}
\end{figure}

\section{Results} \label{sec:results}

\subsection{Temporal Variability}

\subsubsection{GRS 1758\(-\)258}

As discussed above, GRS is a persistent source at hard X-rays/soft gamma-rays.  Figure~\ref{fig:lc_grs} shows the ISGRI light curves on a revolution timescale (\(\sim 2.5-3\) d) in the 30\(-\)50, 50\(-\)100, 100\(-\)300, and 300\(-\)600 keV energy bands.  The average count rate is overplotted as a red dashed line, and a black solid line is overplotted at a count rate of 0 ct/s.  The source was significantly detected in the first four energy bands at 1081.8\(\sigma\), 770.6\(\sigma\), 191.3\(\sigma\), and 5.1\(\sigma\).  There was no detection in the highest energy band (1.1\(\sigma\)).  The coloring for the different revolutions is based on the spectral state analysis performed in Sec.~3.2.1.

%\begin{wrapfigure}{r}{0.5\textwidth}
 \begin{figure}[h!]
  \begin{center}
  \includegraphics[scale=0.9, angle=0,trim = 12mm 15mm 10mm 100mm, clip]{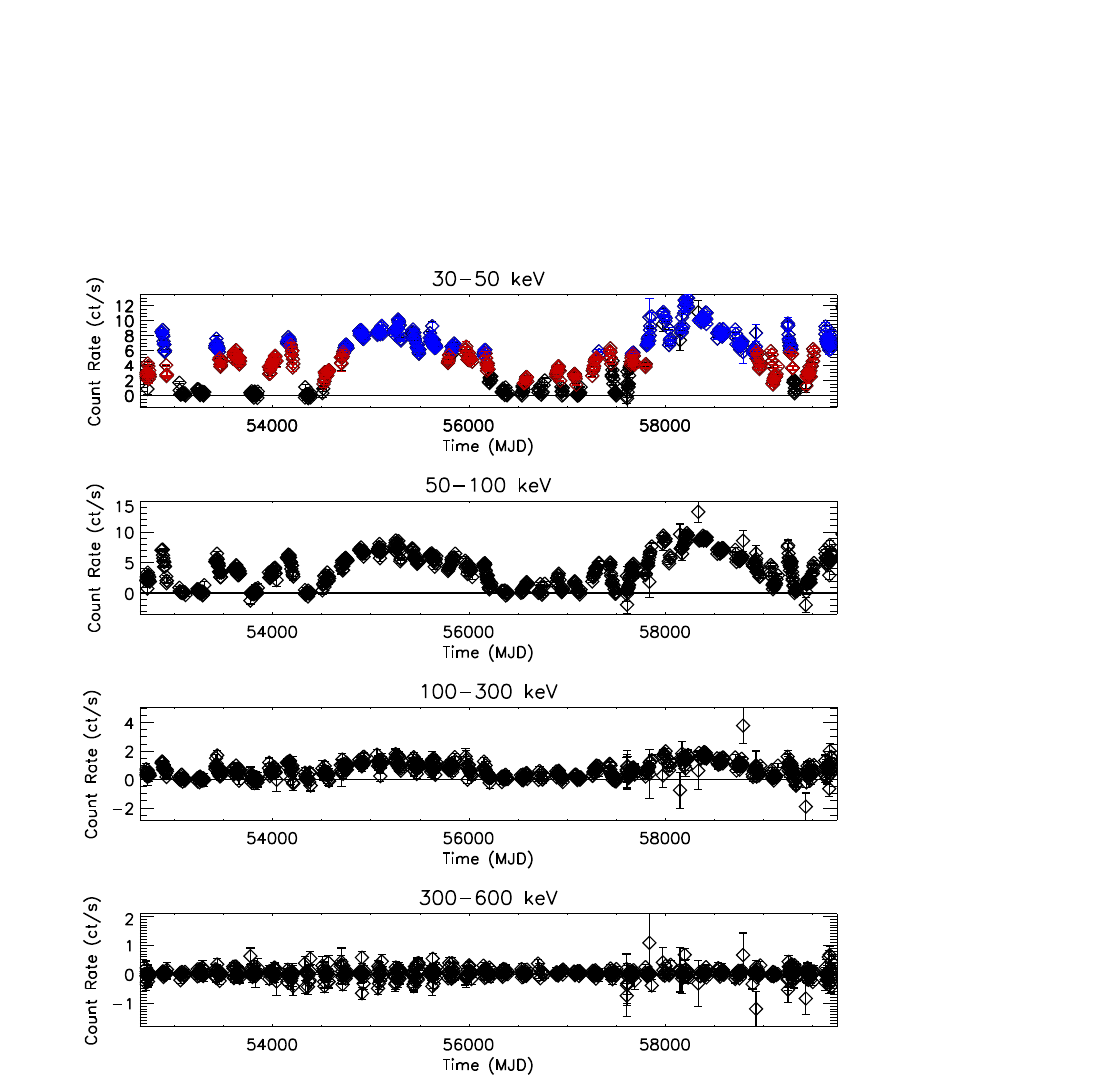  }
    \end{center}
  \caption{ISGRI long-term light curve of 1E in the 30\(-\)50, 50\(-\)100, 100\(-\)300, and 300\(-\)600 keV energy bands.  Black solid line denotes 0 ct/s.  The different colors in the 30\(-\)50 keV light curve correspond to different spectral states discussed in Sec.~3.2.2.} 
\label{fig:lc_1e}
%    \hspace{-60mm}
\end{figure}

\subsubsection{1E 1740.7\(-\)2942}

1E is also a persistent source at hard X-ray/soft gamma-rays, though it displays larger flux variability between observing periods with numerous times where the source count rate is marginally detected, or below the ISGRI detection limit.  The ISGRI revolution timescale light curves are shown in Figure~\ref{fig:lc_1e}.  As in the case of GRS, the 30\(-\)50, 50\(-\)100, 100\(-\)300, and 300\(-\)600 keV energy bands are plotted.  1E was significantly detected in the first four energy bands at 710.3\(\sigma\), 467.3\(\sigma\), 108.0\(\sigma\), and 6.6\(\sigma\), respectively, but it is not detected in the last energy channel (1.7\(\sigma\)).  The coloring for the different revolutions is based on the spectral state analysis performed in Sec.~3.2.2.

\subsection{Spectral Variability}

\subsubsection{GRS 1758\(-\)258}

\begin{figure}
\centering
     \subfloat[][]{\includegraphics[scale=0.7, angle=0,trim = 15mm 10mm 60mm 120mm, clip]{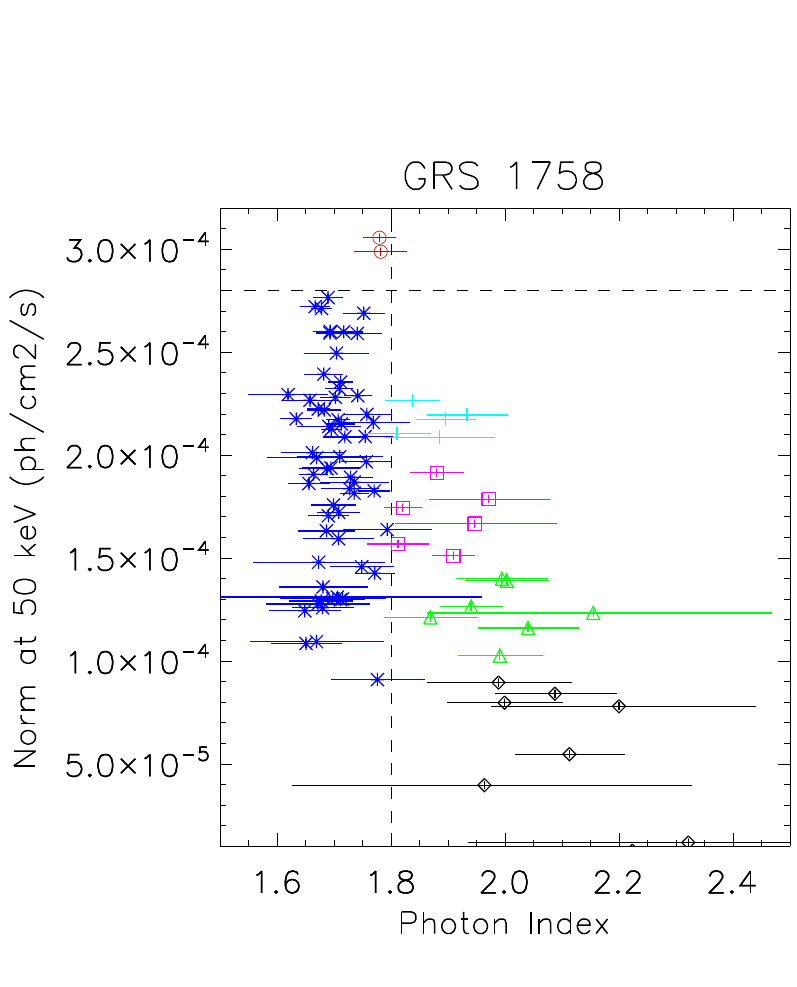}\label{fig:pho_grs}}
     \subfloat[][]{\includegraphics[scale=0.7, angle=0,trim = 15mm 10mm 50mm 120mm, clip]{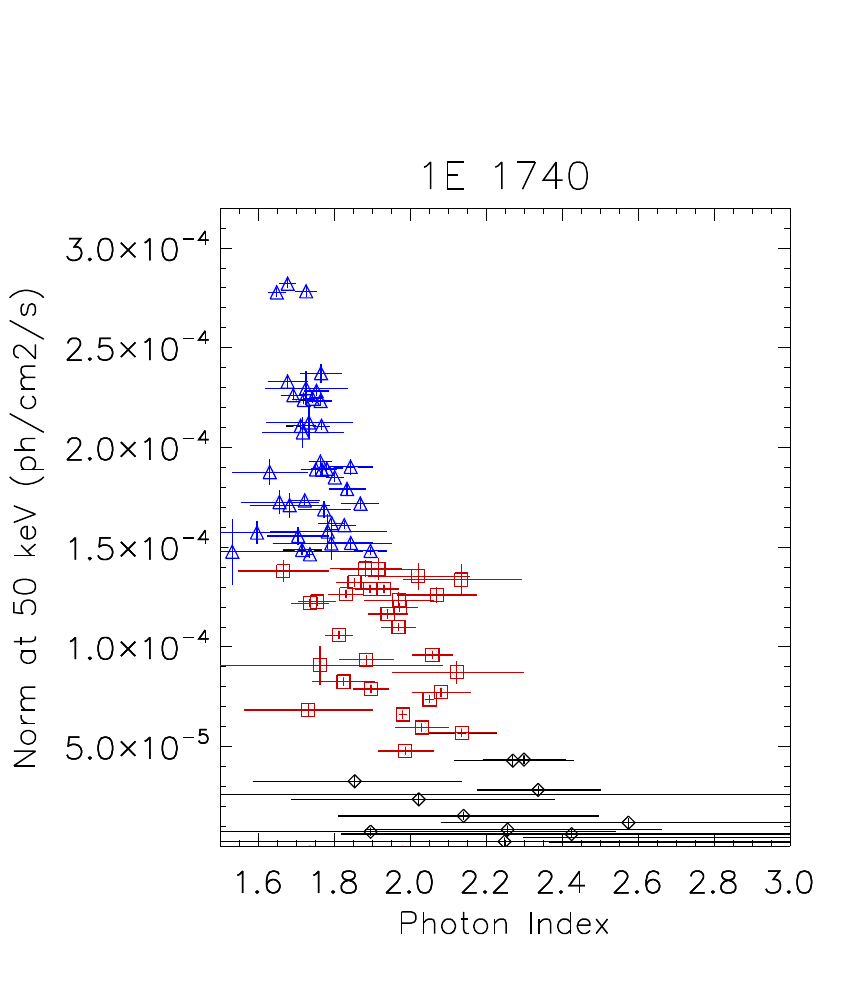}\label{fig:pho_1e}}
     \caption{GRS and 1E photon index versus 50 keV normalization for average spectral groups in panels (a) and (b), respectively.  For GRS, the vertical dashed line at \(\Gamma = 1.8\) marks the separation between the ``hard" and ``soft" groupings.  For 1E, the low flux state corresponds to periods with normalization less than 4.5\(\times 10^{-5}\) ph/cm\(^2\)/s .  The medium flux state corresponds to normalization between 4.5\(\times 10^{-5}\) and 1.45\(\times 10^{-4}\) ph/cm\(^2\)/s.  The high flux state corresponds to normalization between above 1.45\(\times 10^{-4}\) ph/cm\(^2\)/s.}
     \label{steady_state}
\end{figure}

To search for spectral variability, initially data were fit to a power-law model in the 30\(-\)90 keV energy range on a revolution timescale.  However, the photon indexes were poorly constrained.  Thus spectra from revolutions close in count rate and close in time were averaged to maximize statistics when fitting.  The average photon indexes and the normalizations at 50 keV are plotted in Figure~\ref{fig:pho_grs}.  

Though the normalization varies by a factor of roughly 7 (excluding the very low count rate periods), the photon index does not show any clear correlation.  For spectra with \( \Gamma < 1.8\), the photon index is roughly constant at \( \Gamma = 1.7\) for normalizations from \(\sim 8 \times 10^{-5} - 2.8 \times 10^{-4}\) ph/cm\(^2\)/s.  These spectral parameters are shown as blue asterisks in Figure~\ref{fig:pho_grs}.  At a normalization of approximately 3\(\times 10^{-4}\) ph/cm\(^2\)/s, the photon indices have a value of roughly \(1.75 - 1.8\), which are plotted as red circles.  These two spectral groupings are subsequently referred to as the ``hard medium" and ``hard high" states.  

The vertical dashed line at \(\Gamma = 1.8\) separates the ``hard" and ``soft" states.  In contrast to the hard states, the soft states show a clearer relationship between photon index and normalization with \(\Gamma\) decreasing as the normalization increases.  The soft spectra were grouped into four normalization levels: below 1\(\times 10^{-4}\) ph/cm\(^2\)/s (soft low, plotted as black diamonds), 1\(\times 10^{-4}\) to 1.45\(\times 10^{-4}\) ph/cm\(^2\)/s (soft medium 1, plotted as green triangles), 1.45\(\times 10^{-4}\) to 2.05\(\times 10^{-4}\) ph/cm\(^2\)/s (soft medium 2, plotted as magenta squares), and greater than 2.05 \(\times 10^{-4}\) ph/cm\(^2\)/s (soft high, plotted as cyan plus signs).  

The spectra within each spectral grouping were averaged together for six spectra.  First, the spectra were fitted to a power law in the 30\(-\)150 keV energy range.  The fit parameters for each spectrum are listed in Table~\ref{table:spectra}.  Only the soft low spectrum was acceptably described by a power law, which has a best-fit photon index of \(2.06 \pm 0.03\).  The residuals for the other spectra suggest the presence of a high-energy cutoff.  Thus we fitted them using a cutoff power-law model (\texttt{cutoffpl}), and each spectrum is well described, though with significantly different \(\Gamma\) and cutoff energy (\(E_{cut}\)) values.  The soft medium 1, soft medium 2, and soft high spectra have decreasing photon indices from \( \Gamma \sim 1.7\) to 1.4 and decreasing \(E_{cut}\) from \(\sim 240\) keV to 110 keV.  The hard medium and hard high spectra show comparable photon indices (\(1.25 \pm 0.03\) and \(1.36 \pm 0.09\), respectively) and cutoff energies (\(137 \pm 8\) keV and \(120 \pm 16\) keV, respectively).  However, the parameters are not an acceptable for the hard medium spectrum (\(\chi^2/\nu = 51.80/26 = 1.99\)).

\begin{figure}
\centering
    \hspace{-10mm}
     \subfloat[][]{\includegraphics[scale=0.7, angle=0,trim = 15mm 120mm 60mm 20mm, clip]{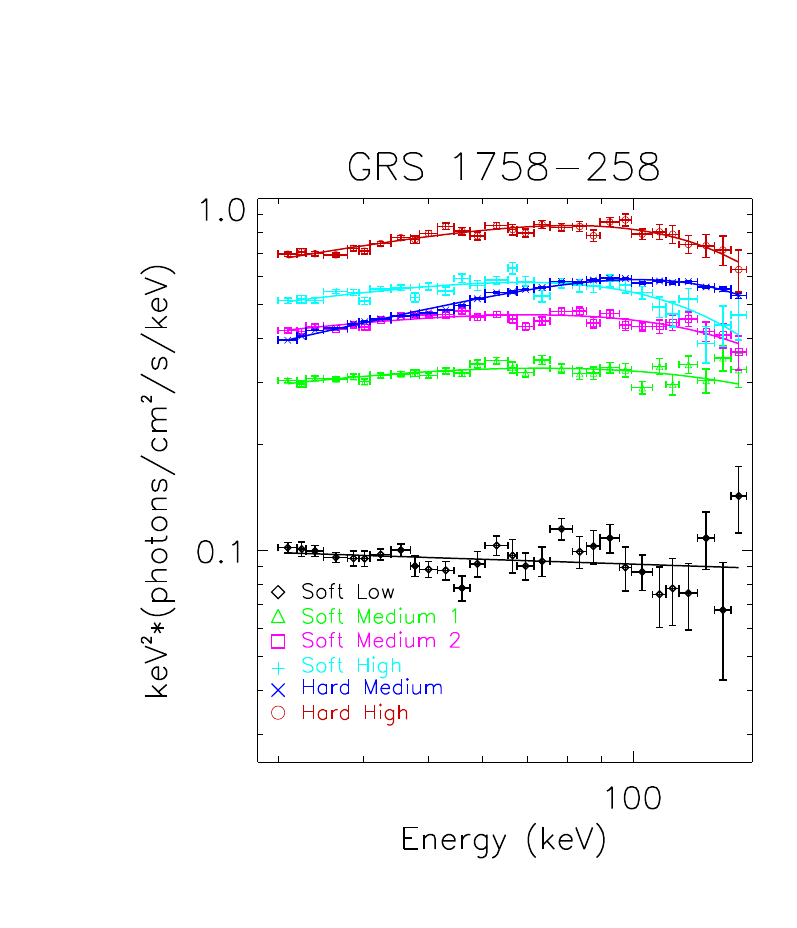}\label{fig:spec_var_grs}}
     \hspace{-10mm}
     \subfloat[][]{\includegraphics[scale=0.7, angle=0,trim = 15mm 120mm 50mm 20mm, clip]{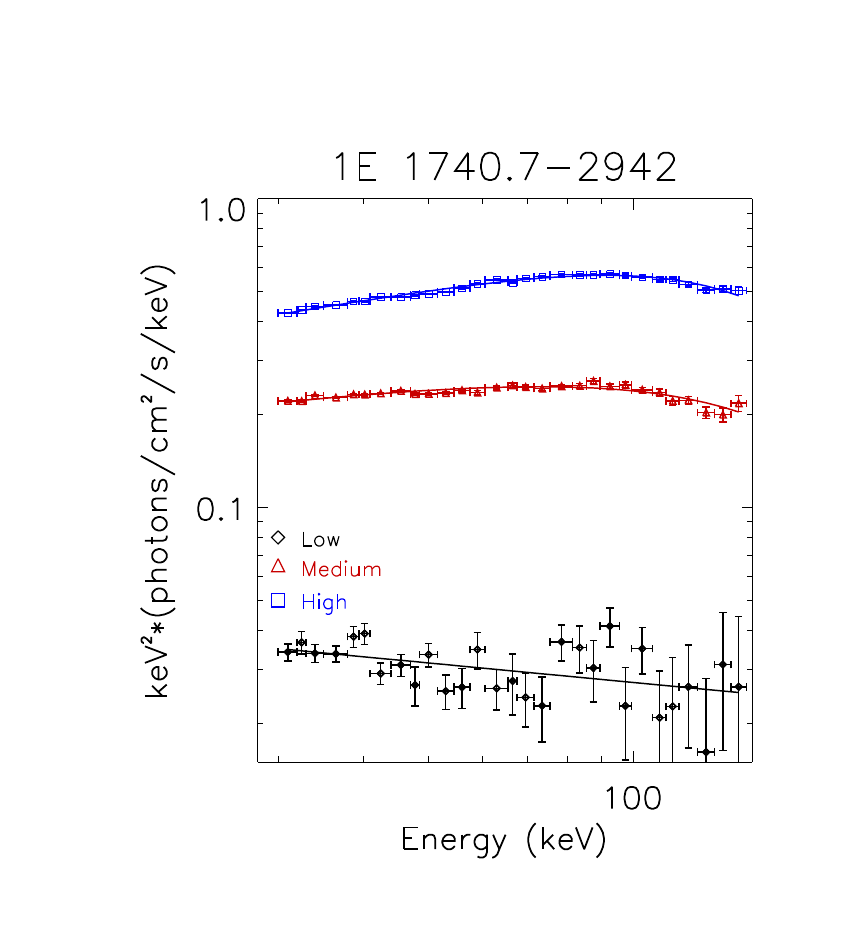}\label{fig:spec_var_1e}}
     \caption{GRS (a) and 1E(b) average state spectra in the 30\(-\)150 keV energy range with \texttt{CompTT} model fits overplotted for each spectrum, except for the low state spectra, which are shown with power-law models.}
     \label{steady_state}
\end{figure}

Subsequently, we fitted all the non-low state spectra to the thermal Comptonization model \texttt{CompTT} assuming a disk accretion geometry.  For each, the photon seed temperature (\(kT_0\)) was fixed to 0.5 keV.  As with the cutoff power-law model, the electron temperature (\(kT_e\)) decreases as flux increases from \(\sim 70\) keV for the soft medium 1 spectrum to 34 keV for the soft high spectrum, though medium 1 spectrum is poorly constrained (\(\pm 68\) keV).  On the other hand, the optical depth (\(\tau\)) increases as flux increases, going from 0.7 to 1.5.  The hard spectra have similar \(kT_e\) values (38.8 and 35, for the medium and high spectra, respectively), but they have significantly different \(\tau\) values (1.73 and 1.6, for the medium and high spectra, respectively).  The average spectra with their \texttt{CompTT} model are plotted in Figure~\ref{fig:spec_var_grs}.

%\begin{figure}
%\begin{subfigure}[r!]{0.6\linewidth}
%\hspace{-10mm}
%\includegraphics[scale=0.7, angle=0,trim = 15mm 10mm 60mm 120mm, clip]{grs1758_spec8_paper_90kev_pho.pdf}
%\caption{\label{fig:pho_grs}}
%\end{subfigure}
%\hfill
%\hspace{-20mm}
%\begin{subfigure}[l!]{0.6\linewidth}
%\includegraphics[scale=0.7, angle=0,trim = 15mm 10mm 50mm 120mm, clip]{1e1740_spec4_paper_90kev_pho_250129.pdf}
%\caption{\label{fig:pho_1e}}
%\end{subfigure}%
%\caption{GRS (a) and 1E (b) average state spectra in the 30\(-\)150 keV energy range with \texttt{CompTT} model fits overplotted for each spectrum, except for the low state spectra, which are shown with power-law models.}

%\caption{GRS and 1E photon index versus 50 keV normalization for average spectral groups in panels (a) and (b), respectively.  For GRS, the vertical dashed line at \(\Gamma = 1.8\) marks the separation between the ``hard" and ``soft" groupings.  For 1E, the low flux state corresponds to periods with normalization less than 4.5\(\times 10^{-5}\) ph/cm\(^2\)/s .  The medium flux state corresponds to normalization between 4.5\(\times 10^{-5}\) and 1.45\(\times 10^{-4}\) ph/cm\(^2\)/s.  The high flux state corresponds to normalization between above 1.45\(\times 10^{-4}\) ph/cm\(^2\)/s.}
%\end{figure}

Due to the large amount of exposure time, and very stable instrumental background level, GRS is significantly detected above 150 keV in the hard medium state.  Thus we extended the spectral fits up to 610 keV.  For the \texttt{CompTT} model \(kT_e\) is \(44.0 \pm 0.8\) keV, larger than the fit up to 150 keV, and \( \tau\) decreases from 1.73 to 1.54 (\( \chi^2 / \nu = 61.90/36 = 1.72 \)).  The fit parameters for this model and subsequent models are listed in Table~\ref{table:joint_spec}.  However, above \( \sim\) 200 keV the residuals are all above the model, suggesting the presence of a possible high-energy component.  Therefore we added a power-law model to fit the spectrum above 200 keV.  The \( \chi^2 / \nu \) improved to \(36.82/34 =1.08 \) with \(kT_e = 36\) keV, \(\tau =\)1.9, and \(\Gamma = \) 1.8.  Thus a high-energy component is needed to acceptably fit the spectrum above 150 keV.  To test the significance of the power-law component, we performed an F-test and found an F-statistic of 17.62 and probability of \( 5.61 \times 10^{-6}\) (\(\sim 4.5 \sigma\)).   The spectrum with the \texttt{CompTT+powerlaw} model is plotted in Figure~\ref{fig:avg_spec} as black diamonds.  The medium and high spectra for 1E are also shown as red squares and blue triangles, respectively.  

Similar high-energy excesses have been observed in other accreting black holes (e.g. GX 339\(-\)4 \citep{johnson1993}, Cyg X-1 \citep{mcconnell2000}, GRS 1915\(+\)105 \citep{zdziarski2001}), which has largely been confirmed with INTEGRAL.  (See \cite{motta2021}).  The origin of the emission is debated.  Polarization results of the high-energy spectrum of Cyg X-1 from INTEGRAL \citep{laurent2011,jourdain2012} suggest that the emission associated with the jet observed at radio wavelengths.  However, the excess can also be explained by a hybrid thermal/non-thermal population of electrons in the corona \citep{coppi1999}.  Thus we fit the spectrum to the \texttt{Eqpair} model.  Following \cite{gierlinski1999}, \cite{delsanto2008}, and \cite{bassi2020}, we left the \( l_h / l_s\) (the ratio of the hard to soft compactness), \(l_{nt} / l_n\) (the fraction of power going to the energetic particles that accelerates the non-thermal particles), and \(\tau_p\) (the Thomson scattering depth) parameters free while \(l_{bb}\) (the soft photon compactness) was fixed to 10 and \(G_{inj}\) (the index of the electron distribution) was fixed to 2.5.  Additionally, \(kT_{bb}\) (the temperature of the inner edge of the accretion disk) was fixed to 0.5 keV, the ionization parameter of the reflector (\(\xi\)) and the reflection fraction were fixed to 0.  The best-fit parameters are \( l_h / l_s = 6.7 \pm 0.1\), \(l_{nt} / l_h = 0.58 \pm 0.07\), and \(\tau_p = 1.1 \pm 0.2\) with \( \chi^2 / \nu = 30.05/35 = 0.86 \).  Thus the model also well describes the data.

%\begin{figure}
%\begin{subfigure}[r!]{0.6\linewidth}
%\hspace{-10mm}
%\includegraphics[scale=0.7, angle=0,trim = 15mm 120mm 60mm 20mm, clip]{grs1758_flx_avg_250115.pdf}
%\caption{\label{fig:spec_var_grs}}
%\end{subfigure}
%\hfill
%\hspace{-35mm}
%\begin{subfigure}[l!]{0.6\linewidth}
%\includegraphics[scale=0.7, angle=0,trim = 15mm 120mm 50mm 20mm, clip]{1e1740_flx_avg_250121.pdf}
%\caption{\label{fig:spe_var_1e}}
%\end{subfigure}%
%\caption{GRS (a) and 1E(b) average state spectra in the 30\(-\)150 keV energy range with \texttt{CompTT} model fits overplotted for each spectrum, except for the low state spectra, which are shown with power-law models.}
%\end{figure}

\subsubsection{1E 1740.7\(-\)2942}

The same search for spectral variability was performed for 1E.  Figure~\ref{fig:pho_1e} shows the photon index and normalization behavior.  In contrast to GRS, 1E exhibits a more straight forward relationship between the 50 keV flux and photon index.  Groupings with normalizations below 4.5\( \times 10^{-5}\) ph/cm\(^2\)/s (low state, plotted as black diamonds) have \(\Gamma \sim 2.2\), though many of the photon indices are poorly constrained.  For normalizations from 4.5\( \times 10^{-5}\) to 1.45\( \times 10^{-4}\) ph/cm\(^2\)/s (medium state, plotted as red squares), groupings have an average \(\Gamma\) of roughly 1.9.  Finally, the high state (\(> 1.45 \times 10^{-4}\) ph/cm\(^2\)/s, plotted as blue triangles) have an average photon index of approximately 1.75.  

As with GRS, we averaged the spectral groupings together and fitted them with a power-law model.  Again, the low flux state is consistent with a power law \(\Gamma = 2.21 \pm 0.07\) while the medium and high flux state spectra are suggestive of a high-energy cutoff.  The fit parameters for all the models are shown in Table~\ref{table:spectra}.  We fitted the medium and high spectra to a cutoff power-law model.  The medium spectrum fit improves, but is still not acceptable (\(\chi^2/\nu = 34.29/26 = 1.32\)) for \(\Gamma = 1.59 \pm 0.04\) and \(E_{cut} = 168 \pm 20\) keV.  Also, the \(\chi^2 / \nu\) for the high state is still unacceptable (35.52/26 = 1.37) with \(\Gamma = 1.32 \pm 0.03\) and \(E_{cut} = 130 \pm 8\) keV.

Next, we fitted both spectra to a \texttt{CompTT} model.  In this case, the model describes the data well \( \chi^2 / \nu = 27.97/26 = 1.04\) for the medium state and \(17.56/26 = 0.68\) for the high state.  The medium spectrum has a \(kT_e\) of 44 keV and \(\tau\) of 1.2 compared to  \(kT_e\) of 38 keV and \(\tau\) of 1.61 for the high spectrum.

The medium and high flux spectra show significant emission above 150 keV.  Thus we again investigated the presence of a high-energy tail by fitting the spectrum up to 610 keV.  For the medium spectrum, the \texttt{CompTT} fit has an increased \(kT_e\) from 44 keV to 57 keV and a decreasing \(\tau\) from 1.2 to 0.9.  The \( \chi^2 / \nu \) is acceptable (1.25), but again the high-energy residuals are all above the model and suggest the presence of an additional spectral component.  The fit parameters are listed for this fit and subsequent fits in Table~\ref{table:joint_spec}.  Including a power-law at high energies with \(\Gamma = 1.3\) reduces the electron temperature (35 keV) and increases the optical depth (1.3) and reduces the \(\chi^2\) to 33.37 for \(\nu =34\) (\( \chi^2/\nu = 0.98\)).  Results of an F-test found an F-statistic of 5.99 and probability of \(0.0059\) (\(\sim 2.8 \sigma\)), and thus the additional component is not significant.

%\begin{figure}[h!]
\begin{wrapfigure}{r}{0.5\textwidth}
  \begin{center}
  \vspace{-9mm}
%  \hspace{-10mm}
  \includegraphics[scale=0.7, angle=0,trim = 35mm 125mm 0mm 50mm, clip]{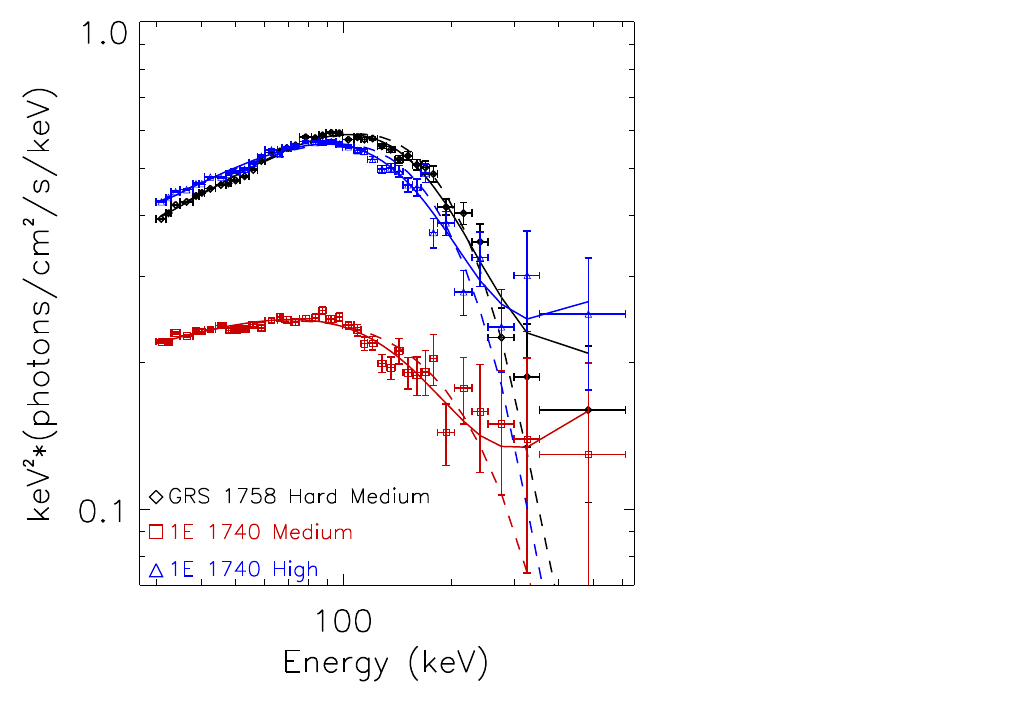}
  \caption{Average state spectra for GRS hard medium (black diamonds) and 1E medium (red squares) and high (blue triangles) in the 30\(-\)610 keV energy range.  The \texttt{CompTT+powerlaw} model is overplotted as a solid line for each while the \texttt{CompTT} model is overplotted as a dashed line.} 
\label{fig:avg_spec}
    \end{center}
%\vspace{7mm}
\end{wrapfigure}

Nonetheless, we fitted the spectrum with the \texttt{Eqpair} model.  The fit parameters are \( l_h / l_s = 3.8 \pm 0.2\), \(l_{nt} / l_h = 0.84 \pm 0.08\), and \(\tau_p = 1.2 \pm 0.1\) with \( \chi^2 / \nu = 30.35/35 = 0.87\).  The \( \tau_p\) values are similar to GRS while the \( l_nt / l_h\) and \( l_h / l_s\) are significantly different (3.8 compared to 6.7 and 0.58 compared to 0.84, respectively).

%\( l_nt / l_h\) and \( \tau_p\) are similar to the values for GRS while the \( l_h / l_s\) is significantly different (3.6 compared to 6.5).  }

We performed the same analysis for the high state.  In this case the \texttt{CompTT} model is a poor fit to the model (\( \chi^2 / \nu = 68.67/36 = 1.91\)) due to the excess at high energies.  Adding a power-law component significantly improves the quality of the fit (\( \chi^2 / \nu = 33.72 / 34 = 0.99\)) with \(kT_e = 34\) keV, \(\tau = 1.7\), and \( \Gamma = 1.5\). An F-test analysis finds an F-statistic of 17.62 and a probability of \(5.61\times 10^{-6}\) (\(\sim 4.5 \sigma\)). An \texttt{Eqpair} model fit also acceptably fits the data (\( \chi^2 / \nu = 31.36 / 35 = 0.90\)) with \( l_h / l_s = 5.9 \pm 0.2\), \(l_{nt} / l_h = 0.87 \pm 0.09\), and \(\tau_p = 1.1 \pm 0.2\).  The \(\tau_p\) values are comparable to GRS 1758, but \(l_{nt} / l_h\) is significantly higher.  The  \( l_h / l_s \) value of GRS is higher than both those for 1E.  As before, the medium and high spectra are also shown as red squares and blue triangles, respectively, in Figure~\ref{fig:avg_spec}.

\begin{table*}
\begin{center}
\caption{Average Spectra Parameters}
 \hspace{-22mm}
\scalebox{0.75}{
\hspace{-24mm}
\begin{tabular}{cccccccccc}
\tableline 
\multicolumn{1}{c}{} & \multicolumn{8}{c}{GRS 1758\(-\)258}  &    \\
\tableline
&  \multicolumn{2}{c}{\texttt{Power-Law}}                  & \multicolumn{3}{c}{\texttt{Cutoff Power-Law}}                                       & \multicolumn{3}{c}{\texttt{CompTT}} & Exp. Time\\
              & \( \Gamma\)       &    \(\chi^2 / \nu\)    &    \( \Gamma\)     &   \( E_{cut}\)         & \(\chi^2 / \nu\)                      &   \(kT_e\)        & \(\tau\)          &     \(\chi^2 / \nu\)  & (Ms) \\
&             &                   &                                             &   ( keV)               &                                       &    (keV)          &                   &                       &  \\
                 %&      &                       &                        &         &                     &            &                     \\
\tableline                   \\
Soft Low      & \(2.06 \pm 0.03\) &  \(35.50/27 = 1.31\)   & \(-\)    & \(-\)   &  \(-\)                 & \(-\)                                 & \(-\)             & \(-\)             & 1.04                       \\
Soft Medium 1 & \(1.93 \pm 0.01\) &  \(43.87/27 = 1.32\)   & \(1.69 \pm 0.06\)  & \(242 \pm 63\)         &  \(28.14/26 = 1.08\)                  & \(70 \pm 68\)    & \(0.7 \pm 0.8\)    & \(28.89/26 = 1.11\) & 1.44 \\
Soft Medium 2 & \(1.97 \pm 0.01\) &  \(78.98/27 = 2.93\)   & \(1.52 \pm 0.08\)  & \(139 \pm 21\)         &  \(28.68/26 = 1.10\)                  & \(48 \pm  8\)    & \(1.1 \pm 0.2\)   & \(30.68/26 = 1.18\) & 0.97 \\
Soft High     & \(1.95 \pm 0.02\)& \(90.58/27 = 3.36\)   & \(1.42 \pm 0.08\)  & \(110 \pm 16\)           &  \(35.18/27 = 1.35\)                  & \(34 \pm  3\)    & \(1.5 \pm 0.1\)    & \(31.72/26 = 1.22\) & 0.45 \\
Hard Medium   & \(1.739 \pm 0.007\)& \(359.14/27 = 13.19\) & \(1.25 \pm 0.03\)  & \(137 \pm  8\)         &  \(51.80/26 = 1.99\)                  & \(38.8 \pm  1.0\)& \(1.73 \pm 0.04\)  & \(21.49/26 = 0.83\) & 20.43\\
Hard High     & \(1.85 \pm 0.01\)  & \(90.58/32 = 3.36\)   & \(1.36 \pm 0.09\)  & \(120 \pm 16\)         &  \(25.38/26 = 0.98\)                  & \(35 \pm  2\)    & \(1.6 \pm 0.1\)    & \(22.51/26 = 0.87\) & 0.35 \\
%Soft Low      & \(2.05 \pm 0.03\) &  \(38.47/32 = 1.20\)   & \(-\)    & \(-\)   &  \(-\)                 & \(-\)                                 & \(-\)             & \(-\)             & 1.04                       \\
%Soft Medium 1 & \(1.95 \pm 0.01\) &  \(52.89/32 = 1.65\)   & \(1.65 \pm 0.06\)  & \(211 \pm 45\)         &  \(26.96/31 = 0.87\)                  & \(62 \pm 24\)    & \(0.8 \pm 0.4\)    & \(27.13/31 = 0.88\) & 1.44 \\
%Soft Medium 2 & \(1.88 \pm 0.02\) &  \(49.16/32 = 1.54\)   & \(1.54 \pm 0.08\)  & \(176 \pm 40\)         &  \(24.53/31 = 0.79\)                  & \(44 \pm  6\)    & \(1.29 \pm 0.2\)   & \(22.16/31 = 0.71\) & 0.44 \\
%Soft High     & \(1.968 \pm 0.002\)& \(98.16/32 = 3.07\)   & \(1.42 \pm 0.08\)  & \(110 \pm 15\)         &  \(32.90/31 = 1.06\)                  & \(36 \pm  3\)    & \(1.4 \pm 0.1\)    & \(31.47/31 = 1.02\) & 0.45 \\
%Hard Medium   & \(1.801 \pm 0.008\)& \(484.44/32 = 15.14\) & \(1.25 \pm 0.03\)  & \(133 \pm  7\)         &  \(32.65/31 = 1.05\)                  & \(43 \pm  1\)    & \(1.55 \pm 0.04\)  & \(20.75/31 = 0.67\) & 18.81\\
%Hard High     & \(1.88 \pm 0.01\)  & \(95.50/32 = 2.98\)   & \(1.38 \pm 0.06\)  & \(126 \pm 17\)         &  \(24.33/31 = 0.78\)                  & \(40 \pm  3\)    & \(1.4 \pm 0.1\)    & \(26.35/31 = 0.85\) & 0.35 \\
\tableline      
\multicolumn{1}{c}{} &  \multicolumn{8}{c}{1E 1740.7\(-\)2942} &  \\
\tableline
& \multicolumn{2}{c}{\texttt{Power-Law}}            & \multicolumn{3}{c}{\texttt{Cutoff Power-Law}}               & \multicolumn{3}{c}{\texttt{CompTT}}  & Exp. Time\\
        & \( \Gamma\)         &    \(\chi^2 / \nu\) &    \( \Gamma\)      &   \( E_{cut}\)  & \(\chi^2 / \nu\)    &   \(kT_e\)      & \(\tau\)         &      \(\chi^2 / \nu\) &  \\
&             &                        &                                  &   ( keV)        &                     &    (keV)        &                                          & \\
\tableline
%Low     & \(2.14 \pm 0.06\)        & \(39.38/32 = 1.23\)                    & \(-\)    & \(-\) & \(-\)                 & \(-\)           & \(-\)            & \(-\)                 &  4.54        \\
%Medium  & \(1.99 \pm 0.01\)  & \(133.15/32 = 4.16\) & \(1.57 \pm 0.05\)  & \(156 \pm 17\)  & \(28.39/31 = 0.92\)  & \(47 \pm 4\)    & \(1.1 \pm 0.1\)  & \(25.11/31 = 0.81\)   & 10.37  \\
%High    & \(1.871 \pm 0.008\)& \(390.49/32 = 12.20\)& \(1.30 \pm 0.03\)  & \(126 \pm 8\)   & \(42.52/31 = 1.37\)  & \(41 \pm 1\)    & \(1.50 \pm 0.05\)& \(34.91/31 =1.13 \)   & 8.88  \\
Low     & \(2.21 \pm 0.07\)    & \(33.33/27 = 1.23\)  & \(-\)              & \(-\)           & \(-\)                & \(-\)           & \(-\)            & \(-\)                 &  4.55  \\
Medium  & \(1.953 \pm 0.009\)  & \(115.49/27 = 4.28\) & \(1.59 \pm 0.04\)  & \(168 \pm 20\)  & \(34.29/26 = 1.32\)  & \(44 \pm 4\)    & \(1.2 \pm 0.1\)  & \(27.97/26 = 1.04\)   & 10.62  \\
High    & \(1.829 \pm 0.007\)  & \(304.72/27 = 11.29\)& \(1.32 \pm 0.03\)  & \(130 \pm 8\)   & \(35.52/26 = 1.37\)  & \(38 \pm 1\)    & \(1.61 \pm 0.05\)& \(17.56/26 =0.68 \)   & 10.83  \\

\tableline
\label{table:spectra}
\end{tabular}
%\end{center}
}

\end{center}
\end{table*}

\section{Discussion} \label{sec:disc}

\subsection{Comparison with Previous Works}

\subsubsection{GRS 1758\(-\)258}

There have been several works studying the long-term behavior of GRS up to roughly a couple hundred keV to compare with our INTEGRAL observations.  GRANAT/SIGMA monitored GRS (and 1E) in the 40\(-\)200 keV energy range from 1990\(-\)1998.  \cite{kuznetsov1999} found cutoff spectral parameters of \( \Gamma = 1.0 \pm 0.3\) and \(E_{cut} = 89^{+40}_{-20}\) keV and \texttt{CompTT} parameters of \(kT_e = 41^{+7}_{-5}\) keV and \(\tau = 1.2 \pm 0.2\).  These values are most similar to the 30\(-\)150 keV hard medium spectrum.  \(\Gamma\) and \(kT_e\) are comparable to the ISGRI values (\(\Gamma = 1.25 \textrm{ and } kT_e = 38.8\) keV), but the \(E_{cut} \textrm{ and } \tau\) values are significantly different (137 keV and 1.73), suggesting that the spectral curvature in the SIGMA spectrum is different from the ISGRI one.

BeppoSAX observed GRS on 10\(-\)11 April, 1997 in the \(0.1-200\) keV energy range.  The continuum spectrum was well described by an absorbed power law with a high-energy cutoff (\texttt{powerlaw*highecut}) with \(\Gamma = 1.65\), \(E_{cut} = 73\) keV, and \(E_{fold} = 180 \) keV \citep{sidoli2002}.  We fitted our hard medium spectrum with the same model without the low-energy absorption in the 30\(-\)210 keV energy range and found comparable parameters (\(\Gamma = 1.60\), \(E_{cut} = 85\) keV, and \(E_{fold} = 185 \) keV).  The authors also fit the spectrum to a \texttt{CompTT} model, but assuming a spherical accretion geometry.  Thus we refitted the hard medium spectrum using the same assumption.  Again the fit resulted in similar parameters (ISGRI: \(kT_e = 42\) keV, \(\tau = 3.8\) and BeppoSAX: \(kT_e = 44\) keV, \(\tau = 3.6\)).

\cite{pottschmidt2008} analyzed the INTEGRAL/SPI and IBIS data from early in the mission (2003\(-\)2007) and reported the presence of a hard tail above \(\sim 300\) keV.  Their best fit \texttt{CompTT+po} model had \(kT_e = 41\) keV, \(\tau = 1.4\), and \(\Gamma = 1.4\), though they found that it is possible to get an equally good fit with \( \Gamma = 2.0\), comparable to the photon index that we report (1.8).  However, we tested our spectrum with \(\Gamma = 1.4\), and the fit is comparable (\(38.56/35 = 1.10\) vs \(36.82/34 = 1.08\)) with \(kT_e = 38\) keV and \(\tau = 1.70\).

\subsubsection{1E 1740.7\(-\)2942}

As mentioned in the previous section, SIGMA observed 1E from 1990\(-\)1998.  Its average spectrum was well fit by a cutoff spectrum with \( \Gamma = 1.2 \pm 0.02\) and \(E_{cut} = 110^{+26}_{-23}\) keV \citep{kuznetsov1999}, similar to the high state spectrum (\( \Gamma = 1.32\), \(E_{cut} = 130\) keV), though with a lower \(\Gamma\).  For the \texttt{CompTT} fit, the \(kT_e\) values are not similar (44 keV for SIGMA and 35 keV for ISGRI) while the optical depths are (1.63 compared to 1.60 for SIGMA and ISGRI, respectively).  Thus it is unclear which spectral state the SIGMA observations correspond to compared to the ISGRI spectrum.

%, but the optical depths are significantly different 1.63 compared to 1.50 for SIGMA and ISGRI, respectively).  Thus it is unclear which spectral state the SIGMA observations correspond to compared to the ISGRI spectrum.  

\cite{bouchet2009} analyzed the ISGRI and INTEGRAL/SPI observations from 2003\(-\)2005 in the 18\(-\)600 keV energy range.  They found a significant excess above 200 keV.  Thus their spectrum more closely matches our high state spectrum, which has a similar 30 keV flux.  Including a power law for the high-energy component to the \texttt{CompTT} model resulted in an acceptable fit with \(kT_e = 27 \pm 2.2\) keV, \(\tau = 1.9 \pm 0.25\), and \(\Gamma = 1.9 \pm 0.1\)  compared to  (\(kT_e = 35 \pm 6\) keV, \(\tau = 1.3 \pm 0.2\), \( \Gamma = 1.3 \pm 1.0\)) for our spectrum.  Fixing \(kT_e\) and \(\Gamma\) to the values from \cite{bouchet2009}, we find \(\tau = 2.56 \pm 0.03\), significantly higher.    

%, which is comparable to our spectrum (\(kT_e = 35 \pm 2\) keV, \(\tau = 1.69 \pm 0.06\), \( \Gamma = 1.5 \pm 1.4\)).

They also tested a scenario where the high-energy emission is due to a second thermal Comptonization component for a \texttt{CompTT+CompTT} model with best-fit parameters \(kT_{e_1} = 29.4 \pm 3.1\) keV, \(\tau_1 = 1.6 \pm 0.1\), \(kT_{e_2} = 100\) keV (fixed), and \( \tau_2 = 2.2 \pm 0.8\).  The seed photon temperature for the second \texttt{CompTT} component was tied to \(kT_{e_1}\).  We fitted our spectrum to this scenario with comparable parameters for the \texttt{CompTT} component (\(kT_{e_1} = 25 \pm 3\) keV and \(\tau = 1.77 \pm 0.06\)), but the optical depth for the second component is significantly different (\(1.0 \pm 0.3\)).

\begin{table*}
\begin{center}
\caption{30\(-\)610 keV Fit Spectral Parameters}
 \hspace{-25mm}
\scalebox{0.70}{
\hspace{-25mm}
\begin{tabular}{cccccccccccc}
\tableline 
\multicolumn{1}{c}{} & \multicolumn{11}{c}{GRS 1758\(-\)258}      \\
\tableline
 & \multicolumn{3}{c}{\texttt{CompTT}} & \multicolumn{4}{c}{\texttt{CompTT+po}} & \multicolumn{4}{c}{\texttt{Eqpair}} \\
\tableline
 & \(kT_e\)   &    \( \tau\) &  \(\chi^2 / \nu\)  & \( kT_e \) & \(\tau\) & \(\Gamma\) & \(\chi^2 / \nu\)  & \(l_h/l_s\)  &  \(l_{nt} / l_h\) & \(\tau_p\) & \(\chi^2 / \nu\) \\
 &   (keV)    &              &                    &   (keV)    &          &            &                   &              &                     &            & \\
\tableline                   
Medium & \(44.0 \pm 0.8\) & \(1.54 \pm 0.08\) & \(61.90/36 = 1.72\) & \(36 \pm 3\) & \(1.9\pm 0.3\) & \(1.8 \pm 0.2\) & \(36.82/34 = 1.08\) & \(6.7 \pm 0.1\) &  \(0.58 \pm 0.07\) & \(1.1 \pm 0.2\) & \(30.05/35 = 0.86\) \\
%Medium & \(46 \pm 1\) & \(1.46 \pm 0.04\) & \(38.73/36 = 1.08\) & \(36 \pm 3\) & \(2.0 \pm 0.3\) & \(1.9 \pm 0.2\) & \(22.03/34 = 0.65\) & \(6.5 \pm 0.4\) &  \(0.56 \pm 0.06\) & \(1.2 \pm 0.5\) & \(19.34/35 = 0.55\) \\
\tableline
  \multicolumn{1}{c}{} & \multicolumn{11}{c}{1E 1740.7\(-\)2942}    \\
\tableline
Medium & \(57 \pm 6\) & \(0.9 \pm 0.1\) & \(45.13/36 = 1.25\) & \(35 \pm 6\) & \(1.3 \pm 0.2\)  & \(1.3 \pm 1.0\) & \(33.37/34 = 0.98\) & \(3.8 \pm 0.2\) &  \(0.84 \pm 0.08\) & \(1.2 \pm 0.1\) & \(30.35/35 = 0.87\) \\
%\textbf{Medium} & \(\)         & \(\)            & \(\)                & \(31 \pm 4\) & \(1.7 \pm 0.2\)  & \(1.9 \) (fixed)& \(30.89/35 = 0.88\) & \(\)            &  \(\)              & \(\)            & \(\)                \\
High   & \(43 \pm 4\) & \(1.44 \pm 0.04\)& \(68.67/36 = 1.91\)& \(34 \pm 4\)  & \(1.7 \pm 0.3\) & \(1.5 \pm 0.8\) & \(33.72/34 = 0.99\) & \(5.9 \pm 0.2\) &  \(0.87 \pm 0.09\) & \(1.1 \pm 0.2\) & \(31.36/35 = 0.90\) \\
%\textbf{High}   & \(\)         & \(\)             & \(\)               & \(30 \pm 2\)  & \(2.2 \pm 0.2\) & \(1.9 \) (fixed) & \(35.33/34 = 1.04\) & \(6.1 \pm 0.2\) &  \(0.86 \pm 0.10\) & \(1.1 \pm 0.2\) & \(40.17/34 = 1.18\) \\
%Medium & \(59 \pm 7\) & \(0.9 \pm 0.2\) & \(34.72/36 = 0.96\) & \(33 \pm 13\) & \(1.4 \pm 0.7\)  & \(1.6 \pm 1.3\) & \(25.52/34 = 0.75\) & \(3.6 \pm 0.2\) &  \(0.58 \pm 0.09\) & \(1.1 \pm 0.1\) & \(25.33/35 = 0.72\) \\
%High  & \(43 \pm 1\) & \(1.45 \pm 0.05\)& \(64.81/36 = 1.80\) & \(33 \pm 5\)  & \(1.7 \pm 0.4\) & \(1.5 \pm 0.9\) & \(35.69/34 = 1.05\) & \(6.1 \pm 0.2\) &  \(0.88 \pm 0.10\) & \(1.0 \pm 0.2\) & \(35.16/35 = 1.00\) \\
\tableline
\end{tabular}
}

\label{table:joint_spec}
\end{center}
\end{table*}

\subsection{Origin of the High-energy Emission}

As mentioned above, the physical mechanism that produces the high-energy excess, or hard tail, remains unclear.  A possible explanation is that the photons are produced by a hybrid thermal/non-thermal Comptonizing plasma \citep{coppi1999}.  This model has been used to explain the hard tails in Cyg X-1 in the soft state \citep{gierlinski1999,cangemi2021}, GX339-4 \citep{delsanto2008}, GRS 1716-249 \citep{bassi2020}, among other sources.  

Another suggested interpretation is that the hard tail is due to a second thermal Comptonizing component.  \cite{bouchet2009} applied this interpretation to 1E, as seen in the previous section.

A third possibility is that the excess is non-thermal emission related to the radio jet.  Observations since the early 2000s have found steady, compact jets in radio observations of numerous black hole binaries in the hard state \citep{fender2004}.  During the soft state, the jet is either not present or extremely faint \citep{russell2019,maccarone2020}.  During the hard to soft state transition, the jet becomes unstable and flares are seen at radio wavelengths (\cite{fender2006} and references within.  Polarization detections by INTEGRAL of Cyg X-1 \citep{laurent2011,jourdain2012}, MAXI J1348\(-\)630 \citep{cangemi2023}, and Swift J1727.8\(-\)1613 \citep{bouchet2024} support this interpretation, at least in some sources.  

GRS and 1E radio observations have long reported the existence of persistent jets (e.g. \cite{rodriguez1992,mirabel1999,marti2017,tetarenko2020}).  A study of the spectral states in GRS with RXTE/PCA and VLA during 2001-2003 found behavior consistent with the standard hard/soft paradigm \citet{soria2011} presented above.  For 1E multi-wavelength analysis during July 2007 show only upper limits from ATCA radio observations during our soft state \citep{dickey2023}, suggesting that 1E also follows the expected behavior.  

In the jet interpretation, the soft gamma-ray photons are produced at the base of the jet \citep{zdziarski2012}.  In contrast, the emission is connected to the corona in the hybrid thermal/non-thermal plasma and second thermal Compontionzing component interpretations.  In the hybrid thermal/non-thermal model, the thermal and non-thermal electrons come from the accretion flow \citep{poutanen2014}.  The non-thermal electrons are possibly produced by shock acceleration \citep{fragile2008,das2009,henisey2012} or magnetic reconnection \citep{ding2010,riquelme2012,hoshino2013}. Thus any changes in the accretion flow that effect the population of thermal electrons are expected to also effect the non-thermal electron population, resulting in a correlation between the two components.  Thus a possible way to differentiate between the corona and jet interpretations is to investigate the correlations between the hard X-rays and the soft gamma-rays.

\begin{figure}
\centering
    \hspace{-10mm}
     \subfloat[][]{\includegraphics[scale=0.6, angle=0,trim = 15mm 10mm 30mm 140mm, clip]{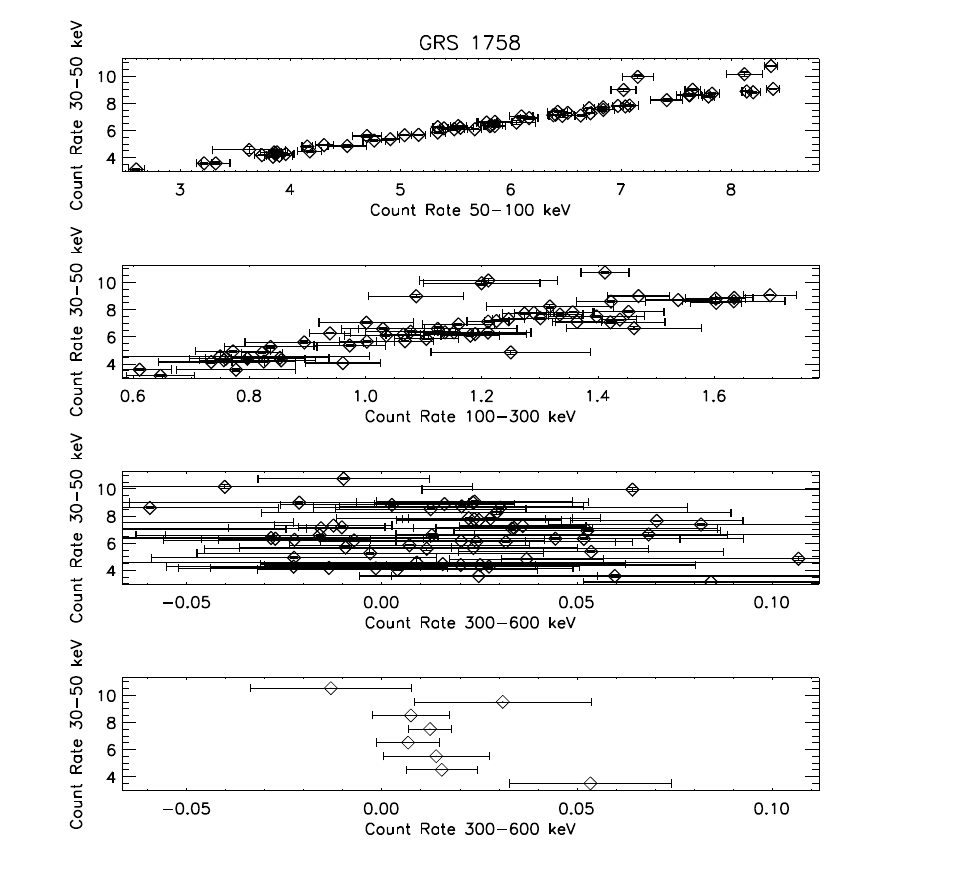}\label{fig:corr_grs}}
     \hspace{-10mm}
     \subfloat[][]{\includegraphics[scale=0.6, angle=0,trim = 15mm 10mm 50mm 140mm, clip]{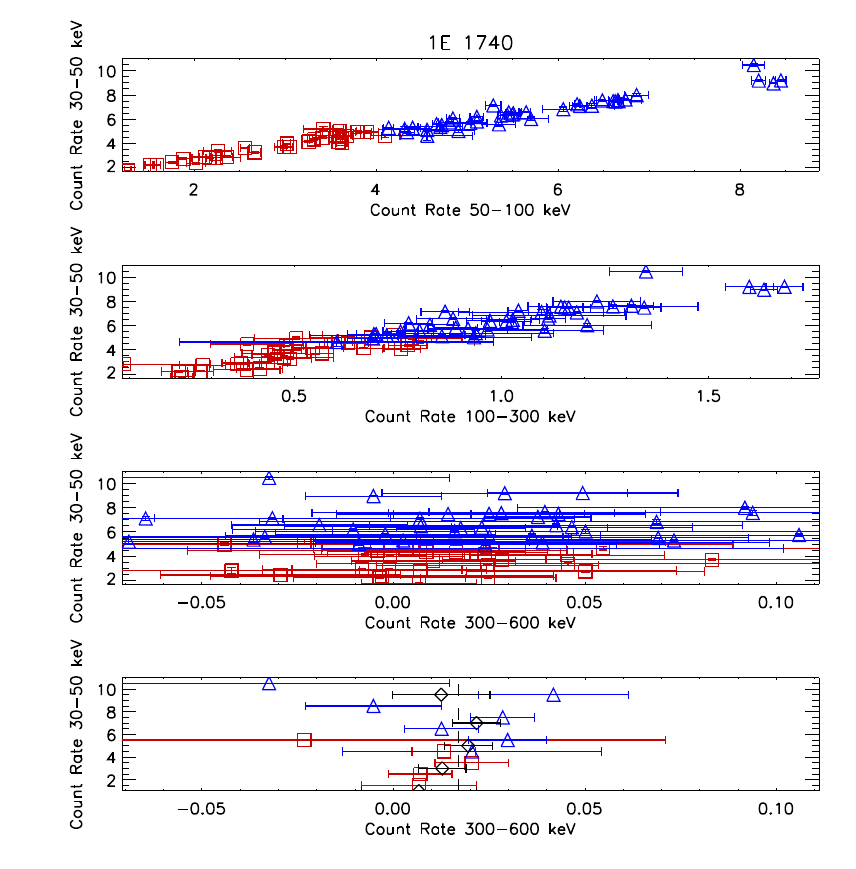}\label{fig:corr_1e}}
     \caption{The \(50-100\) keV (\textit{first}), \(100-300\) keV (\textit{second}), and \(300-600\) (\textit{third and fourth}) keV count rates versus \(30-50\) keV count rate for the GRS hard medium state (a) and the 1E medium and high states (b), which are plotted as red squares and blue triangles, respectively.  The bottom panel shows the \(300-600\) versus \(30-50\) keV count rates with the data rebinned in 1 ct/s bins and the 2 counts/s bins plotted as black diamonds with the best-fit constant value overplotted as a black dotted line.}
     \label{steady_state}
\end{figure}

Therefore, we looked for correlations in count rates between four broad energy bands in the GRS hard medium state and the 1E medium and high states. Figure~\ref{fig:corr_grs} shows the \(30-50\) keV count rate versus the \(50-100\) keV count rate (first panel), \(100-300\) keV (second panel), and \(300-600\) keV (third and fourth panels).  The bottoms panels show the same data as the third panel, but the data were rebinned to 1 counts/sec bins.  Figure~\ref{fig:corr_1e} shows the same plots for 1E with the medium state plotted as red squares and the high state plotted as blue triangles.  For each panel, we calculated the Spearman correlation coefficient and the standard deviation from a null correlation.

In the case of GRS, the Spearman coefficients (and standard deviations) are 0.98 (-7.63), 0.86 (-6.65), and -0.07 (0.58) for the 50\(-\)100, 100\(-\)300, and 300\(-\)600 keV energy bands, respectively.  There are strong correlations between the 30\(-\)50 keV band count rate and the count rates in the next two higher energy bands.  However, there is no strong correlation between the 30\(-\)50 keV and 300\(-\)600 keV count rates.  The lack of correlation may be due to large errors in the \(300-600\) keV count rates so we rebinned the data in 1 counts/sec bins based on the 30\(-\)50 keV count rate.  Even with higher statistics, there is still no significant correlation between the two energy bands with a coefficient of -0.57 (1.51), though there is a weak anti-correlation between the two energy bands, which is biased by the highest and lowest count rate points.  If they are excluded, the coefficient (and standard deviation) are 0.02 (-0.06).  

%\begin{figure}
%\begin{subfigure}[r!]{0.6\linewidth}
\hspace{-10mm}

For 1E, the medium state correlations (and standard deviations) are 0.92 (-4.96), 0.84 (-4.51), and -0.04 (0.23) for the 50\(-\)100, 100\(-\)300, and 300\(-\)600 keV energy bands, respectively.  For the high state the values are 0.95 (-5.84), 0.85 (-5.27), and 0.21 (-1.27) for the same energy bands as the medium state.  Thus in both states the there are strong correlations between the 30\(-\)50 keV count rates and the count rates in the next two higher bands, but no significant correlation between with the 300\(-\)600 keV count rates.  Again, the 300\(-\)600 keV count rates are low so we rebinned the data as was done for GRS and calculated the correlations (and standard deviations).  There is still no strong correlation with coefficients of -0.10 (0.20) and -0.29 (0.70) for the medium and high states, respectively.  The scatter between the points was still large so we combined the data in 2 counts/s bins and across spectral states (shown in the bottom panel with black diamonds).  In this case, there is a weak positive correlation of 0.40 (-0.80).  As with GRS, the lowest and highest count-rate values heavily influence the results.  Excluding the lowest point, the correlation is -0.20 (0.35) while excluding the highest point the corelation is 1.00 (-1.73).  However, the \(300-600\) count rates are consistent with a constant value of \(0.017 \pm 0.003\) ct/s (\( \chi^2 /\nu = 1.79/4 = 0.45\)), which is plotted as a black dashed line in the bottom panel of Figure~\ref{fig:corr_1e}.  

The strong correlation between the count rates below 300 keV suggest that those photons originate from the same mechanism.  This is consistent with the spectral fitting, which finds that thermal Comptonization dominates up to \(\sim 200\) keV.  However, a lack of strong correlation between the hard X-rays (30\(-\)50 keV) and the soft gamma-rays (300\(-\)600 keV) suggests that the high-energy photons arise from a different mechanism.  These results then favor the jet interpretation over the corona interpretation for the origin of the high-energy excess in GRS.  For 1E, the interpretation is less clear.  The results when rebinning in 2 counts/s intervals are suggestive of a correlation when including all the points (0.40), but the correlation coefficient is heavily depenent on the lowest count rate value, and the count rates are consistent with a constant value.   Detections of polarization in the sources could help determine the origin of the hard tails.  However, the 300\(-\)600 keV fluxes for the sources are quite low compared to previous source with polarization detections from IBIS or SPI.  The GRS hard medium flux is roughly 50 mCrab compared to the Crab spectrum reported in \cite{jourdain2020}.  The 1E medium and high fluxes are approximately 33 and 60 mCrab, respectively.  The hard-tail fluxes are roughly an order of magnitude lower than those of Cyg X-1 \citep{laurent2011}, making the detection of polarization difficult \citep{motta2021}.  

\section{Conclusion} \label{sec:con}
INTEGRAL's extensive monitoring of GRS and 1E for over 20 years has enabled an in-depth study of the spectral behavior of the two sources from hard X-rays to soft gamma-rays, and we investigated their characteristics up to 600 keV.  In the case of GRS, we found it to be predominately in a spectral state with a Comptonized spectrum that does not vary with flux (\(\sim 20/25\) Ms).  For comparatively brief periods during \textbf{our} observations, the source displayed a flux dependent spectral behavior where the 30\(-\)90 keV photon index decreased as the 50 keV flux increased.  In contrast, 1E exhibited a more straight forward spectral-flux relationship where the 30\(-\)90 keV spectrum hardens are the the 50 keV flux increases, though the spectrum is relatively stable in the flux ranges: \(< 4.5 \times 10^{-5}\) ph/cm\(^{2}\)/s, \(4.5 \times 10^{-5} - 1.45 \times 10^{-4}\) ph/cm\(^2\)/s, \(> 1.45 \times 10^{-4}\) ph/cm\(^{2}\)/s.

Three of the spectra showed significant emission up to \(\sim 600\) keV (GRS hard medium, 1E medium and high).  Subsequently, we fitted the 30\(-\)610 keV energy range to a \texttt{CompTT} model.  The residuals of each spectrum showed an excess above roughly 200 keV.  For the GRS hard medium and 1E high state spectra the \(\chi^2/\nu\) values are unacceptable, while the fit to the 1E medium state is still acceptable.  The residuals in each fit showed and excess above \(\sim 200\) keV.  Thus we fitted the spectra with a \texttt{CompTT+powerlaw} model.  The \(\chi^2 /\nu\) improves in each case, but the improvement was not significant for the 1E medium state (\(\sim 2.8 \sigma\)). For the GRS hard medium and 1E high state, the fits become acceptable (\( \chi^2 / \nu =\) 1.08 and \(= 0.99\), respectively).  

%For the 1E high state spectrum the \(\chi^2/\nu\) is unacceptable while the other fits are still acceptable.  The residuals in each fit showed an excess above \(\sim 200\) keV.  Thus we fitted the spectra with a \texttt{CompTT+powerlaw} model.  While the \(\chi^2 /\nu\) improves in each case, an F-test found that the improvement was not significant for the 1E medium state (\(\sim 2.8 \sigma\)) and marginal in the GRS hard medium spectrum (\(\sim 4 \sigma\)).  For the 1E high state, the fit becomes acceptable (\( \chi^2 / \nu = 1.05\)).  

As the nature of the high-energy excess remains unclear, we tested a scenario where those photons are from a hybrid thermal/non-thermal corona using the \texttt{Eqpair} model.  In each fit, the fraction of power devoted to energetic particle accelerating the non-thermal particles is \( > 0.58\) and statistically significant (\( > 8 \sigma\)), supporting the presence of a high-energy excess in each spectrum.

%\( > 0.5\) and statistically significant (\( > 6 \sigma\)), supporting the presence of a high-energy excess in each spectrum.  

However, the hard tails may also be associated with the radio jet, as has been reported for Cyg X-1, MAXI J1348\(-\)630, and Swift J1728.8\(-\)1613, based on polarization measurements.  To differentiate between the two scenarios we investigated the energy dependent correlations for each.  We found that the 30\(-\)50 keV count rates are strongly correlated with the count rates below 300 keV.  However, above 300 keV the behavior changes.  For GRS and the 1E medium and hard states there are weak anti-correlations (-0.57, -0.10, and -0.29, respectively) between the 30\(-\)50 keV and \(300-600\) keV energy bands when rebinning the data for higher statistics.  

The lack of a positive correlation between the 30\(-\)50 keV and 300\(-\)600 keV count rates suggest that photons for the two energy bands are produced in different locations, or by different physical processes,  which would be in tension with the hybrid thermal/non-thermal Comptonization scenario.  Thus these results seem to support a jet origin for the high-energy photons at least in the case of GRS, but due to the low count rate in the 300\(-\)600 keV energy band, the anti-correlations and lack of correlation are not significant enough to  exclude the possibility of a positive correlation.  For 1E, the interpretation is more complicated due to the large scatter in the \(300-600\) keV count rates, even when rebinned.  Further rebinning results in a positive, but weak, correlation, though with few bins and combining different spectral states.  However, the 300\(-\)600 keV count rate is consistent with a constant value, supporting a jet origin interpretation, but not excluding a corona origin interpretation.

\acknowledgments

The authors thank the referee for their comments and suggestions to improve the paper.  The authors thank the Italian Space Agency for the financial support under the “INTEGRAL ASI-INAF” agreement n◦ 2019-35-HH.0. The research leading to these results has received funding from the European Union’s Horizon 2020 Programme under the AHEAD2020 project (grant agreement n. 871158). Based on observations with INTEGRAL, an ESA project with instruments and science data centre funded by ESA member states (especially the PI countries: Denmark, France, Germany, Italy, Switzerland, Spain) and with the participation of Russia and the USA. 


\vspace{5mm}

\end{document}